\documentclass[aps,prl,twocolumn,nofootinbib]{revtex4-1} 

\usepackage[utf8,latin1]{inputenc}   	  				 
\usepackage[T1]{fontenc}          	    			     
\usepackage[british]{babel}        					     
\usepackage{mathpazo}  			   	      				 
\usepackage[dvipsnames,x11names]{xcolor} 				 
\usepackage[babel]{microtype}                            
\usepackage{amsmath,amssymb,amsthm,bm,amsfonts,bbm}      
\usepackage{mathtools}									 
\usepackage{soul}
\usepackage[colorlinks=true,citecolor=green,linkcolor=Purple,urlcolor=Magenta]{hyperref}     

\newcommand{\ket}[1]{\vert#1\rangle}
\newcommand{\bra}[1]{\langle#1\vert}

\newcommand{\ketbra}[2]{\vert #1 \rangle \langle #2 \vert}

\newcommand{\ie}{\textit{i.e.}}
\newcommand{\eg}{\textit{e.g.}}

\DeclarePairedDelimiter{\floor}{\lfloor}{\rfloor}

\makeatletter 
\newcommand{\doublewidetilde}[1]{{%
  \mathpalette\double@widetilde{#1}%
}}
\newcommand{\double@widetilde}[2]{%
  \sbox\z@{$\m@th#1\widetilde{#2}$}%
  \ht\z@=.9\ht\z@
  \widetilde{\box\z@}%
}

\makeatother

\newcommand{\C}{\mathfrak{C}}

\newcommand{\blue}[1]{\textcolor{blue}{#1}}

\newtheorem{theorem}{Theorem}

\newcommand{\todai}{Department of Physics, Graduate School of Science, The University of Tokyo, Hongo 7-3-1, Bunkyo-ku, Tokyo 113-0033, Japan}

\begin{document}

\title{Reversing Unknown Quantum Transformations: \\
 Universal Quantum Circuit for Inverting General Unitary Operations} 

\author{ Marco T\'{u}lio Quintino }
\affiliation{\todai}

\author{   Qingxiuxiong Dong }
\affiliation{\todai}

\author{   Atsushi Shimbo}
\affiliation{\todai}

\author{ Akihito Soeda }
\affiliation{\todai}

\author{ Mio Murao}
\affiliation{\todai}

\date{15th of April 2020}


\begin{abstract}
Given a quantum gate implementing a $d$-dimensional unitary operation $U_d$, without any specific description but $d$, and permitted to use $k$ times, we present a universal probabilistic heralded quantum circuit that implements the {exact} inverse $U_d^{-1}$, whose failure probability decays, exponentially in $k$.  {The protocol employs an adaptive strategy, proven necessary for the exponential performance. It requires $k\geq d-1$, proven necessary for exact implementation of $U_d^{-1}$ with quantum circuits. Moreover, even when quantum circuits with indefinite causal order are allowed, $k\geq d-1$ uses are required.}  We {then} present  a finite set of {linear and positive semidefinite } constraints characterizing universal unitary inversion protocols and formulate a convex optimization problem whose solution is the maximum success probability for given $k$ and $d$.  The optimal values are computed using semidefinite programming solvers for {$k\leq 3$ when $d=2$ and $k\leq 2$ for $d=3$}. {With this numerical approach we show for the first time that indefinite causal order circuits provide an advantage over causally ordered ones in a task involving multiple uses of the same unitary operation.} 
\end{abstract}



\maketitle

\begin{figure}  
	\begin{center}
	\includegraphics[scale=.22]{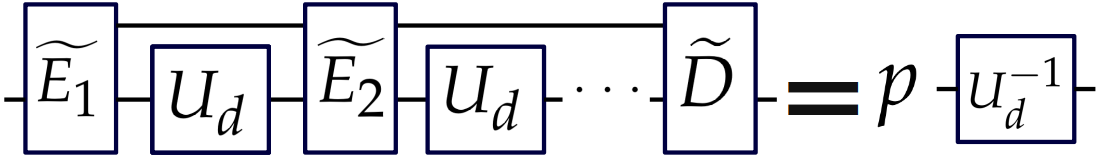} 
	\end{center}
	\caption{{A general quantum circuit that transforms multiple uses of a general $d$-dimensional unitary $U_d$ into its} inverse $U_d^{-1}$ with {the} heralded success probability of {$p$.  The circuit elements between $U_d$'s indicate that the circuit represents an adaptive strategy.  For our adaptive protocol (Theorem\,\ref{theo:seq}), $p=1-\Big(1-\frac{1}{d^2}\Big){\big.}^{\floor{\frac{k+1}{d}}}$ for $k$ uses of $U_d$.}} \label{fig:main}
\end{figure}

Reversible operations in quantum mechanics {are given by unitary operations \cite{cariello12,wolfbook_coro}. } Consider the following problem: a quantum physicist receives a physical apparatus that is guaranteed to perform some qudit unitary operation $U_d$, \eg, a quantum {oracle} \cite{deutsch92,grover96}. Apart from its dimension denoted by $d$, no additional information about this unitary is provided. Is it possible to implement the action of the inverse operation $U_d^{-1}$ without initially knowing the matrix description of $U_d$? A simple strategy to solve this problem is to perform process tomography \cite{chuang96}, obtain a matrix representation of $U_d$, find the inverse matrix that represents $U_d^{-1}${,} and then {decompose the inverse matrix in terms of elementary quantum gates}. {This tomographic approach  may be very inefficient when compared to other protocols and can never be made exact for a finite number of uses of $U_d$.}  Is there a {universal} way for obtaining the inverse $U_d^{-1}$ with finite uses of $U_d$ without any errors? {We answer these questions by explicitly constructing a quantum circuit, see Fig.\,\ref{fig:main}.}

Non-universal protocols to obtain the inverse of known quantum operations {have been analyzed previously,} \cite{barnum00} but since these previous methods depend on the particular operation to be inverted, they cannot be applied to the universal/unknown case considered in this {Letter}.
Universal protocols are considered in  Ref.\,\cite{chiribella16}, where the authors calculate the best expected fidelity of obtaining the inverse operation $U_d^{-1}$ with a single use of a general $U_d$ in a non-exact deterministic scheme. Motivated by refocusing quantum systems in NMR, Ref.\,\cite{sardharwalla16} presents a probabilistic approximated method to invert unitary operations on closed systems. 
While analy{z}ing the ``unitary learning problem'' \cite{vidal01,sasaki02,gammelmark09,bisio10}, Ref.\,\cite{sedlak18} {finds} a probabilistic exact protocol to transform $k$ uses of a general qubit unitary $U_2$ into its inverse $U_2^{-1}$ in a ``store and retrieve'' framework. {Motivated by controling time-dynamics of quantum system, Ref.\,\cite{navascues17,trillo19} presents protocols for universal unitary inversion in a with a limited control over the target system. As such universal unitary inversion protocols can be used to cancel an undesired quantum unitary evolution. 
 Additionally, there are applications for quantum communication without shared reference frames, see Ref.\,\cite{bartlett08} and Sec.\,IV of Ref.\,\cite{bisio10}.}

{In this Letter, we focus on exact protocols, \ie, with unit fidelity, that may fail with some {constant} probability, but when successful, they signal their success (see Fig.\,\ref{fig:main}).}
We present a quantum circuit whose failure probability decays exponentially in $k$.
{The circuit represents an adaptive strategy, proven necessary for the exponential performance.  It requires $k\geq d-1$, also proven necessary for exact implementations of $U_d^{-1}$ with quantum circuits. Moreover, even when quantum circuits with indefinite causal order are allowed, $k\geq d-1$ uses are required.}
We {then present a finite set of linear and positive semidefinite  constraints to} formulate the problem of finding the {optimal} quantum circuit {in} terms of semidefinite programming (SDP). With this SDP approach we show that some quantum circuits with indefinite causal order have an advantage over {all conventionally} ordered circuit{s}.
	%
	%



\begin{figure}
\begin{center}
\includegraphics[scale=.24]{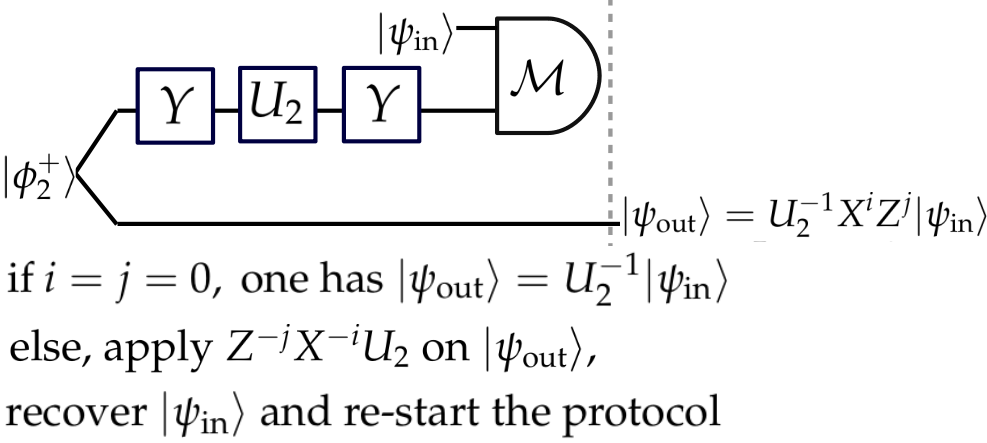} 
\end{center}
\caption{{A quantum circuit to transform} a qubit unitary operation $U_2$ into its inverse $U^{-1}_2$. {First, we prepare the two-qubit maximally entangled state $\ket{\phi^+_2}$ and apply on the upper qubit the gate sequence $Y U_2 Y$, where $Y$ is the Pauli Y operator, which is equivalent to applying $U_2^*$, the complex conjugate of $U_2$.  We then perform a Bell measurement $\mathcal{M}$ on the upper} qubit of $U_2^*\otimes I \ket{\phi_2^+}$ and {on} the input-state $\ket{\psi_\text{in}}$. {After the measurement, $\ket{\psi_\text{in}}$} is transformed into the $U^{-1}_2X^i_2Z^{{j}}_2\ket{\psi_\text{in}}$ with probability $1/4$. When {$i \neq 0$ or $j \neq 0$}, we make an extra use of $U_2$ to recover {$\ket{\psi_\text{in}}$} and repeat the protocol. 
}
\label{fig:qubit_case}
\end{figure}

\textit{{For qubit unitary operations}: -- }  %
	{We start our analysis from qubit unitary operations. Quantum unitary operations can be mathematically represented by linear unitary operators, that is, under a unitary operation, a quantum state $\rho$ evolves as $U \rho U^\dagger$, where $U$ is a linear operator and $U^\dagger$ its adjoint. In this Letter the symbol $U$ may refer to a physical unitary operation or a mathematical linear operator. Its distinction should be clear from the context.} 
{The inverse of a general unitary operator $U_d$ is given by its adjoint operator, \ie, $U_d^{-1}=U_d^\dagger$.  The adjoint of a unitary operator $U_2$ can also be obtained by conjugate transposing the operator with respect to the computational basis \ie, $U_2^\dagger = \big(U_2^T\big)^*$ where $^T$ and $^*$ stands for the transposition and complex conjugation in the computational basis respectively.}
	Our quantum circuit consists in two main parts{, one complex conjugating $U_2$ to $U_2^*$ and the other transposing to $U_2^T$.  Our goal is to implement $U_2^{-1}$ on an input-qubit in an arbitrary (possibly unknown) state $\ket{\psi_\text{in}}$, \ie, to produce a qubit in state $U_2^{-1}\ket{\psi_\text{in}}$.  See Fig.\,\ref{fig:qubit_case} for a pictorial representation.}  
	
	To achieve the complex conjugation, {we first notice that for any dimension $d$ and any real number $\phi$, the unitary operators $U_d$ and $e^{i \phi}U_d$ represent the same physical operation.  Without loss of generality, we can assume that the unitary operator $U_d$ representing a reversible physical operations belongs to $SU(d)$, the group of unitary matrices with determinant one. Then it is enough to perform the Pauli $Y$ operation before and after $U_2$, since $YU_2Y=U_2^*$ for all $U_2\in SU(2)$}. Hence, {the} qubit unitary complex conjugation {is deterministic and exact} \cite{chiribella16,miyazaki17}. 
{The unitary transposition protocol first prepares the maximally entangled state $\ket{\phi_2^+}:=\frac{1}{\sqrt{2}}{(}\ket{00}+\ket{11}{)}$.  We apply $U_2$ on the first qubit of $\ket{\phi_2^+}$ to obtain $U_2\otimes I \ket{\phi_2^+}$, which is mathematically equivalent to $I \otimes U_2^T \ket{\phi_2^+}$.}
	We {then} exploit the gate teleportation scheme \cite{bennett93,gottesman99}, a protocol that ``teleports'' the action of some unitary operation{, which in our case would be $U_2^T$ on $I \otimes U_2^T \ket{\phi_2{^+}}$, to the input-qubit.  This requires} a Bell measurement {$\mathcal{M}$} on the first qubit of $U_2\otimes I \ket{\phi_2{^+}}$ and {the input-qubit}.  {$\mathcal{M}$} consists in projecting the system onto the basis $\{(X^iZ^j)^\dagger \otimes \ket{\phi_2^+}\}$, where $X$ and $Z$ are Pauli operators and $i,j\in\{0,1\}$ are the outcomes of the measurement. Direct calculations show that with probability $1/4$, the state $\ket{\psi_\text{in}}$ is transformed to $U_2^T X^i Z^{{j}} \ket{\psi_\text{in}}$, which is equal to {$U_2^{T}\ket{\psi_\text{in}}$} when $i=j=0$.  {By} concatenating the complex conjugation and the transposition protocols{,} we obtain a circuit that transforms a single use of an arbitrary $U_2$ into $\left(U_2^*\right){}^T   X^i Z^{{j}}  =U_2^{-1}  X^i Z^{{j}} $, which is equal to $U_2^{-1}$ with probability $1/4$. 

	We now consider protocols that make $k$ uses of $U_2$ to obtain $U_2^{-1}$ based on the single-use one described above.
	First note that when this single-use protocol fails, \ie, $j\neq0$ or $i\neq0$, {the protocol ends in} the state $U_2^{-1}  X^i Z^{{j}} \ket{\psi_\text{in}}${.  The initial input-state is ``lost'' as it is.}  In order to {recover} $\ket{\psi_\text{in}}$, we make another use of $U_2$ to apply the operation ${Z^{-j}X^{-i}}U_2$ on the output-state $U_2^{-1} X^i Z^{{j}} \ket{\psi_\text{in}}${.} After this extra use of $U_2$, we have $\ket{\psi_\text{in}}$ {and can} then re-start the {single-use protocol}.  Direct calculation shows that the probability of failing this protocol is $p_\text{F}=\frac{3}{4}^{\floor{\frac{k+1}{2}}}${, where $\floor{x}$ denotes the largest integer not greater than $x$.}

\begin{figure}  
\begin{center}
\includegraphics[scale=.21]{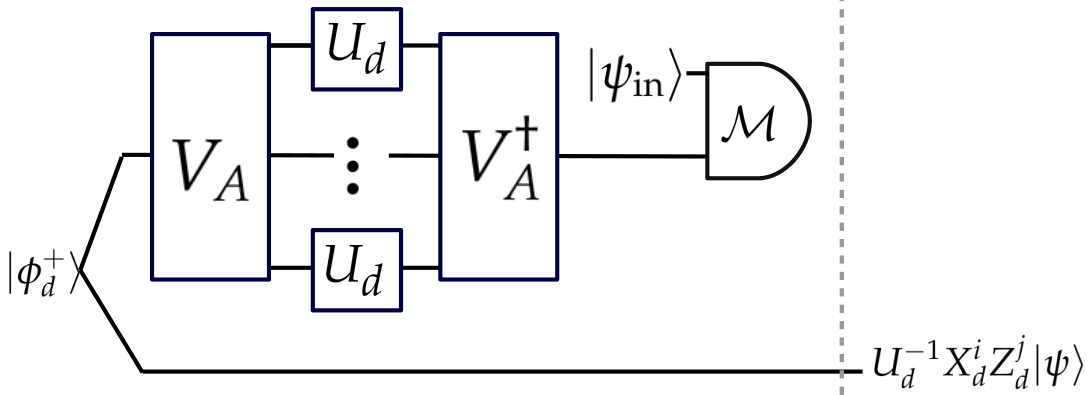} 
\end{center}
\caption{{A quantum circuit that transforms $k=d-1$ uses of a general unitary $U_d$ into its inverse.   It requires a two-qudit maximally entangled state $\ket{\phi^+_d}$ and a generalized Bell measurement $\mathcal{M}$.  The unitary complex conjugation exploits the relation $V_A U_d^{\otimes {d-1}} V_A^\dagger = U_d^*$, where ${V_{A}}$ is defined in the main text.}}\label{fig:fig2}
\end{figure}

\textit{{For general unitary operations}: -- }  %
{Our universal unitary inversion for $d > 2$ also combines the universal transposition and complex conjugation, extending the corresponding subroutines for qubit unitary operations to general $U_d$ {(Fig.\,\ref{fig:fig2} {represents a single round}).}  The transposition is obtained by replacing $\ket{\phi^+_2}$} with the qudit maximally entangled state $\ket{\phi_d^+}:= {\frac{1}{\sqrt{d}}}\sum_{i=0}^{d-1} \ket{ii}$. Later, we apply the operation $U_d$ on the first qudit of $\ket{\phi_d^+}$ to obtain $U_d\otimes I \ket{\phi_d^+}$. 
{The Bell measurement is to generalized to a projective measurement in the basis}  $\{(X_d^iZ_d^j)^\dagger \otimes I \ket{\phi_d^+}\}$, where $X_d^i:=\sum_{l=0}^{d-1} \ketbra{l\oplus i}{l}$ and $Z_d^j:=\sum_{l=0}^{d-1} \omega^{jl} \ketbra{l}{l}$ are the shift and clock operators, respectively, $\omega:= e^{\frac{2\pi \sqrt{-1}}{d}}$, $l \oplus i$ denotes $l+i$ modulo $d$, and $i,j$ are integers ranging from $0$ to $d-1$. Straightforward calculation shows that with probability $1/d^2$, the second {qudit} of $U_d\otimes I \ket{\phi_d^+}$ is transformed into $U_d^T X_d^i Z_d^{{j}} \ket{\psi_\text{in}}$.
	
	For complex conjugation{,} the analogy with qubit case is not as straightforward. {Reference}~ \cite{miyazaki17} presents a deterministic quantum circuit for universal unitary complex conjugation {using $d-1$ of $U_d$.} Let $V_A:\mathbb{C}^d\to{\big(\mathbb{C}^d\big){}^{\otimes d-1}}$ be the isometry mapping qudits into its $(d-1)$ anti-symmetric qudit subspace{, \ie, $V_A := \sum_{\vec{k}} \frac{\epsilon_{\vec{k}}}{\sqrt{(d-1)!}} \ket{k_2,\ldots,k_{d}} \bra{k_1}$,
where $\vec{k} \in \{0,\ldots,d-1\}^{d-1} $ and $\epsilon_{\vec{k}}$ is the antisymmetric tensor of rank $d$.}

It is shown in Theorem 2 of Ref.\,\cite{miyazaki17}) that, for any $U_d\in SU(d)$ one has $V_A^\dagger U_d^{\otimes^{d-1}} V_A = U_d^*$, hence, deterministic unitary conjugation can be obtained simply by performing $V_A$ before using $U_d^{\otimes {(d-1)}}$ and performing $V_A^\dagger$ afterwords.  {We may summarize the foregoing analysis as follows.}

\begin{theorem}\label{theo:seq}
{Given $k$ uses of a unitary operation $ U_d$, its inverse operation can be implemented by a universal circuit with the success probability}
$p_\text{S}=1-\Big(1-\frac{1}{d^2}\Big){\big.}^{\floor{\frac{k+1}{d}}}${.}
\end{theorem}

\begin{figure}  
	  \begin{center}
	\includegraphics[scale=.2]{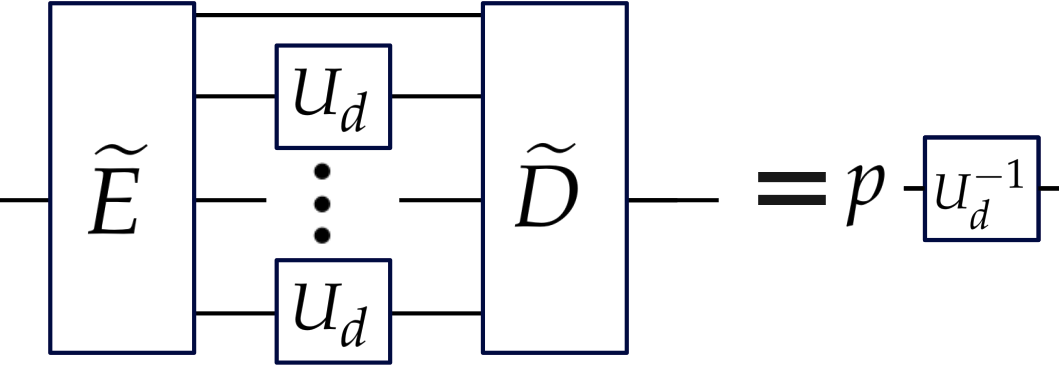} 
	\end{center}
			\caption{Pictorial representation of a parallel{ized} circuit where all the uses of the unitary operations are made simultaneously.} \label{fig:parallel}
	\end{figure}

\textit{{On the exponential scaling of success probability}: --}
{The unitary inversion protocol introduces the measurements $\mathcal{M}$.  Depending on the outcomes of the measurements, each ``round'' either ends in the success ($i=j=0$) or enters the re-starting sequence ($i \neq 0$ or $j \neq 0$).  For each $\mathcal{M}$ introduced, only $d-1$ uses of $U_d$ are applied.  Hence, to increase the success probability at the exponential scaling, \ie, implying larger $k$, the number of rounds increases approximately linearly in $k$.}
{Circuits with parallelized uses of $U_d$, or ``parallelized circuits'', for short, utilize the $k$ uses of operations $U_d$ without any circuit elements introduced in between, or equivalently, call for} a single use of  $U_d^{\otimes {k}}$ (see Fig.\,\ref{fig:parallel} for a pictorial illustration). Parallel{ized} circuits differ from general {adaptive} ones in terms of length and may decrease the time required to complete the full circuit in practical implementations. 
{Nevertheless, the higher concurrency of parallelized circuits is achieved at the expense of the loss of the exponential performance of the adaptive circuits.}	
	\begin{theorem}\label{theo:par_bound}
		Any universal probabilistic heralded parallel{ized} quantum circuit transforming $k$ uses of a $d$-dimension{al} unitary operation $ U_d$ into its 				inverse $ U_d^{-1}$ with a constant {success} probability {$p_\text{P}$ independent of $U_d$} must respect $p_\text{P}\leq 1- {\frac{d+1}{k+d^2-1}}$.
	\end{theorem}
	
The proof begins by assuming that a universal unitary inversion protocol with the success probability $p_\text{P}$ is achieved by a parallelized circuit.  If we run the complex conjugation protocol using $U_d^{-1}$, then this functions as a unitary transposition protocol, since $\Big(U_d^{-1}\Big){\Big.}^* = U_d^T$.  Moreover, this new protocol remains a parallelized circuit as the complex conjugation protocol is parallelized.  %
		The success probability $q$ of this transposition protocol is given by $q = \big(p_\text{P}\big){}^{d-1}$.  In Ref.\,\cite{PRA}, we prove that, within the parallelized circuits, the maximum success probability $q_{\rm{opt}}$ for unitary transposition with $k$ uses of $U_d$ is $q_{\rm{opt}} = 1 - \frac{d^2-1}{k+d^2-1}$, which by definition satisfies $q \leq q_{\rm{opt}}$. Hence, the success probability of any parallel unitary inverse protocol must respect 
	\begin{align}
		p_\text{P} &\leq \left( 1- \left(\frac{d^2-1}{k+d^2-1}\right) \right)^{\frac{1}{d-1}} \\
			   &\leq 1 - \frac{d^2-1}{(d-1)(k+d^2-1)} \label{eq:2}\\
			   &= 1- \frac{d+1}{k+d^2-1},		
	\end{align}
where the Bernoulli inequality\footnote{For real numbers $0 \leq r\leq 1$ and $x\geq -1$ it holds that ${(1+x)^r\leq 1+rx}$ \cite{Bernoulli}.} is used to obtain Eq.\,\eqref{eq:2}.
	We remark that by exploiting similar methods, in Ref.\,\cite{PRA} we prove that $p_\text{P}\leq \frac{d+1}{k+d+1}$, bound which is slightly tighter than the one presented in this letter.
	
{\textit{On the $k\geq d-1$ uses: --}} Our quantum circuit for universal unitary inversion requires at least $d-1$ uses of the un{i}tary $U_d$ {to run the first round.  Less than $d-1$, the protocol cannot start, hence the success probability is zero.  This} is actually required by any protocol transforming a general unitary into its inverse {by an arbitrarily small but nonzero success probability,} showing there is no room for improvement {of} Theorem \ref{theo:seq} in {the} number of uses.
Moreover, this fundamental constraint on {the} number of uses holds even if one considers circuits that make use of the unitaries $U_d$ in an indefinite causal order \cite{chiribella09,oreshkov11,araujo14}.
	\begin{theorem}\label{theo:inv_non}
		Any universal probabilistic heralded quantum circuit (including protocols without definite causal order) transforming $k<d-1$ uses of a $d$-dimension unitary operation $ U_d$ into its inverse $ U_d^{-1}$ with a success probability $p_\text{S}$ that does not depend on $U_d$ necessarily has $p_\text{S}=0$, \ie, null success probability.
	\end{theorem}
{The proof begins by assuming that there exists a unitary inversion protocol using less than $d-1$ instances of $U_d$ with a nonzero success probability.  We combine this inversion protocol with the probabilistic unitary transposition protocol based on gate teleportation.  This achieves a heralded probabilistic (and exact) protocol of universal unitary conjugation, since $\big(U_d^{-1}\big){}^T = U^*_d$, using less than $d-1$ uses of $U_d$.  However, we show in Ref.\,\cite{PRA} that \textit{any} a heralded probabilistic and exact protocol of universal unitary conjugation requires at least $d-1$.  Thus, the unitary inversion protocol using less than $d-1$ instances of $U_d$ leads to a contradiction, hence forbidden.}


\textit{Numerical approach: --}
Our discussion above has been analytical and constructive, thus leading to an analytical expression of the success probability for given $k$ and $d$.  On the other hand, it is not a straightforward task to obtain improved circuits with a higher success probability.  To find the optimal success probability, we must first identify the set of circuits which inverts any $U_d$ in an exact and heralded probabilistic manner.
		
We now present a {finite set of linear and positive semidefinite constraints} characterizing the desired set.
Reference \cite{chiribella07} (see also Ref.\,\cite{kretschmann05}) provides a finite set of linear constraints which are necessary and sufficient conditions for a positive matrix $C$ to represent a deterministic quantum circuit.  More specifically, $C$ corresponds to a \textit{quantum comb}, an incomplete circuit where some gate elements of the circuit are unspecified, thus unexecutable \textit{per se}.  It becomes executable by providing the missing elements as ``inputs'', which in our case would be $U_d$'s.  A heralded probabilistic comb $S$ is represented by a positive matrix satisfying $S \leq C$.
Let $\C(U_{d}):=d (I_d\otimes U_d )\ketbra{\phi_d^+}{\phi_d^+} (I_d\otimes U_d^\dagger)$ be the Choi operator \cite{jamiolkowski72,choi75} {of $U_d$}, and $A*B$ be the link product between two operators $A$ and $B$ \cite{chiribella07}.
If $S$ inverts $k$ uses of $U_d$ with probability $p$, then this is equivalent to {
$S*\C \left( U_{d} \right){}^{\otimes k} = p \; \C\left(\left( U_{d} \right){}^{-1}\right)$.
For $S$ to be a \textit{universal} inversion, then this equation must hold for all $U_d \in SU(d)$.}

{We reduce the number of constraints to finite by observing that $L_{\text{span}} := \text{span} \left\{ \C \left( U_{d} \right) {}^{\otimes k}  \big| U_d \in SU(d)  \right\}$ is a finite linear space.  Thus, there exists a finite subset $\left\{U^{(i)}_{d}\right\}{\big.}_i$ of $SU(d)$ such that 
	$\left\{ \C \left( U^{(i)}_{d} \right){\big.}^{\otimes k} \right\}_i$ form a basis of $L_{\text{span}}$.  Such $U^{(i)}_{d}$ can be found by sampling for a sufficiently large (but finite) number of unitaries $U^{(i)}_{d}$ over the Haar measure, and succeeds, in practice, with the unit probability.}
Now, the maximum success probability $p_\text{S}$ is the solution of
	\begin{equation}  \begin{split} \label{eq:SDP}
						 &\text{max } p_\text{S}  \\ 
		\text{s.t. }  \quad & S*\C \left( U^{(i)}_{d} \right){\big.}^{\otimes k} = p_\text{S} \; \C\left(\left( U^{(i)}_{d} \right) {\big.}^{-1}\right) \; \forall i;   \\ 	
	     &  S \geq0, \;  \quad  C\geq S,  \;  \quad C \; \text{: quantum comb},\\ 
	\end{split}  \end{equation}  
thus is an instance of semidefinite	programming (SDP) with $p_\text{S}$, $S$, and $C$ as the SDP variables. {In Ref.\,\cite{PRA} we discuss this SDP approach in details.}

In Ref.\,\cite{chiribella09}{,} the authors propose a quantum circuit model where the operations may not respect a definite causal order.
{T}he program represented in Eq.\,\eqref{eq:SDP} is flexible enough to consider such indefinite causal order circuits. {T}he operator $C$ must respect a set of linear constraints of a process {matrix}. Such constraints can be found in Ref.\,\cite{oreshkov11,araujo15,araujo16}{,} which we discuss in {detail} in Ref.\,\cite{PRA}. 

 We have implemented the semidefinite program described in Eq.\,\eqref{eq:SDP} with the MATLAB package cvx \cite{cvx} with several SDP solvers \cite{MOSEK,SEDUMI,SDPT3} and summari{z}e {relevant} results in Table~\ref{table:SDP}.  {From Table~\ref{table:SDP} we see that when $d=2$ and $k=2$ or $k=3$ indefinite causal order quantum circuits  strictly outperform any standard ordered one. Previously, indefinite causal order circuits have proven to be useful for tasks whose input quantum operations are not unitary \cite{chiribella09,oreshkov11,ebler18} or consist of distinct unitary operations \cite{araujo14,feix15}. Our example is the first to show that indefinite causal order circuits are useful even when the input quantum operations are restricted to\ multiple uses of the same unitary operation.}

{We emphasize that the SDP solvers also return a numerical description of $S$, hence when considering ordered circuits (quantum combs) one can obtain a gate sequence to realize the corresponding optimal inversion by following the steps of Ref.\,\cite{chiribella07}.}

\begin{table} 
\begin{tabular}{|c|c|c|c|}
\hline 
$d=2$ & Parallel{ized} & {Adaptive} & Indefinite causal order\\
\hline 
$k=1$ & \blue{$\frac{1}{4}=0.25$}  &\blue{$\frac{1}{4}=0.25$}  & \blue{$\frac{1}{4}=0.25$}   \\ 
\hline 
$k=2$ & \blue{$\frac{2}{5}=0.4$}  & {$0.4286\approx \frac{3}{7}$} & $0.4444\approx\frac{4}{9}$  \\ 
\hline 
$k=3$ & \blue{$\frac{1}{2}=0.5$}  & $0.7500\approx\frac{3}{4}$ & 0.9416  \\ 
\hline 
%
%
\hline 
$d=3$  & Parallel{ized} & {Adaptive} & Indefinite causal order\\
\hline 
$k=1$ & \blue{$0$}  & \phantom{aaaa} \blue{$0$} \phantom{aaaa}  & \blue{$0$}  \\ 
\hline 
$k=2$ & {$\frac{1}{9}\approx0.1111$}  & $0.1111\approx \frac{1}{9}$ & $0.1111\approx \frac{1}{9}$  \\ 
\hline 
\end{tabular} 
\caption{{Maximum} success probabilities for {universally inverting} $k$ uses of $U_d${, by parallelized quantum comb (Parallelized), general quantum comb (Adaptive), and circuits with indefinite causal orders (Indefinite causal order)}. Values in blue {are analytical} and {in black} via numeri{c}al SDP optimi{z}ation.}\label{table:SDP}
\end{table}


\textit{Conclusions: -- } %
		We have presented a probabilistic heralded universal quantum circuit that makes $k$ uses of an arbitrary (possibly unknown) $d$-dimensional unitary quantum operation $U_d$ to exactly implement its inverse $U_d^{-1}$. The probability of failure of our main protocol decreases exponentially {in $k$}. We then proved that that the probability of failure of any parallel circuit decays {at most  linearly} in $k${.} We have also shown any circuit (including the ones with indefinite causal order) requires at least $k\geq d-1$ uses of the unitary $U_d$, ensuring that our protocol makes the minimal number of uses to have a nonzero success probability.
		We also provided an SDP characterization of heralded probabilistic quantum combs that achieve exact and universal unitary inversion.  This SDP approach allowed us to find quantum circuits that attain the maximum success probability in parallel, adaptive, and indefinite causal order circuits for {$k\leq 3$ when $d=2$ and $k\leq 2$ for $d=3$}.  We find that there are circuits with indefinite causal order that have a greater success probability than any causally ordered ones. {We have then provided the first task where indefinite causal order circuits are proved to overcome any causally ordered ones even when the input quantum operations are restricted to\ multiple uses of the same unitary operation.
		
		 {From a practical perspective, our protocols rely on repetitions of fixed a sub-circuit  (\eg, Fig.\,\ref{fig:qubit_case}) which does not require adjustments when a higher success probability is required, or when different unitary input-operations are considered.   We hope that our constructions pave the way for novel applications and experimental realisations.}


\textit{Acknowledgements: --} %
 We are indebted to M. Ara{\'{u}}jo, J. Bavaresco, D. Gross, and P. Guerin for valuable discussions.  
 This work was supported by MEXT Quantum Leap Flagship Program (MEXT Q-LEAP) Grant Number JPMXS0118069605, Japan Society for the Promotion of Science (JSPS) by KAKENHI grant No. 15H01677, 16F16769, 16H01050, 17H01694, 18H04286, 18K13467 and ALPS.}

All our code is available in Ref.\,\cite{MTQ_github_unitary} and can be freely used, edited, and distributed under the MIT license \cite{MIT_license}.

\bibliographystyle{linksen.bst}
\bibliography{mtqbib.bib}

\end{document}